# Role of Inter-site Hybrid Interactions in Itinerant Ferromagnetism

**Piyush Dua[1], Sunil Panwar[2] and Ishwar Singh[1]**

[1] Physics Department, Indian Institute of Technology Roorkee, Roorkee – 247 667, INDIA

[2] Department of Basic Sciences, College of Engineering and Technology, Gurukul Kangri University, Haridwar – 249 404, INDIA

We study role of inter-site hybrid interactions in deciding ferromagnetic state in the itinerant electron (narrow band) systems like some of the transition metals. We have considered Hubbard like tight binding model alongwith exchange and hybrid interactions. All interactions have been treated within mean-field approximation. It is found here that hybrid interactions play significant role at zero as well as infinite temperatures in deciding on-set of ferromagnetic state. We have studied variation of effective mass of up and down spin electrons in ferromagnetic state as a function of temperature. Also studied are magnetic susceptibility, optical conductivity and magnetoresistance as a function of temperature.

## I. Introduction

Recently it has been suggested [1] that the onset of ferromagnetism in itinerant electron solids need not be due to the relative shift of the positions of majority and minority spin bands as a result of large intra-atomic Coulomb interactions. Rather the inter-atomic exchange interactions may play crucial roles for the onset of ferromagnetic state. It has been argued that the ferromagnetic state may be realized with inter-atomic exchange interactions alone. The onset of ferromagnetism, in this case, is manifested through the narrowing and widening of, respectively, (majority) up and (minority) down spin electrons. As a result of inter-site exchange interactions, the up spin band is narrowed down and, therefore, is majority filled. Thereby the (kinetic) energy of these electrons is reduced.

In the usual (Stoner) model of itinerant ferromagnetism [2-4], the (large) intra-atomic Coulomb repulsion of electrons gives rise to the difference in band energy between (majority) up and (minority) down spin electrons, when following criterion is fulfilled,

$$U\,\rho\,(\varepsilon_F) > 1 \tag{1}$$

Under this condition, the energy difference between up and down-spin electrons, in the simplest model, is k-independent and is given by

$$\varepsilon_{k\downarrow} - \varepsilon_{k\uparrow} = U(n_\uparrow - n_\downarrow) = U\,M \tag{2}$$

In these equations U is the on-site Coulomb repulsion, $\rho(\varepsilon_F)$ is the density of states at Fermi energy and M is the magnetization per site.

This Stoner effect is overshadowed by the narrowing and widening effect due to inter-atomic exchange interactions [5-9]. In this work, we study the role of (nearest neighbor) inter-site hybrid interactions $\left\langle ii\left|\frac{1}{r}\right|ij\right\rangle$ alongwith inter-site exchange interactions $\left\langle ij\left|\frac{1}{r}\right|ij\right\rangle$. These interactions shall be included in the nondegenerate tight binding Hubbard Hamiltonian.

It may be noted [2] that for d-electrons in transition metals hybrid interactions $\left\langle ii\left|\frac{1}{r}\right|ij\right\rangle$ have got larger magnitude than the exchange interactions $\left\langle ij\left|\frac{1}{r}\right|ij\right\rangle$.

We use Green's function equation of motion method at finite temperature to study role of hybrid interactions in deciding ferromagnetic state. We study variation of effective mass of (majority) up and (minority) down spin electrons in ferromagnetic state as a function of temperature for various values of interaction parameters. Also we study magnetic susceptibility, optical conductivity and magnetoresistance. In the following section (II) we mention the model Hamiltonian and the Green's function treatment within mean field approximation. In section III, we study para/ferro phase diagram at T = 0 K. In section IV, we study and show results of variation of effective mass, magnetic moment, magnetic susceptibility, magneto resistance and optical weight as a function of temperature. In section V, we have discussions and conclusions.

## II. Model Hamiltonian and Green's Function Treatment

We consider non-degenerate tight-binding extended Hubbard model including the nearest-neighbor direct, exchange, pair-hopping and hybrid interactions.

$$H = -t\sum_{<ij>\sigma}(c^{+}_{i\sigma}c_{j\sigma} + H.C.) + U\sum_{i} n_{i\uparrow}n_{i\downarrow} + V\sum_{<ij>} n_{i}n_{j} + J\sum_{<ij>\sigma\sigma'} c^{+}_{i\sigma}c^{+}_{j\sigma'}c_{i\sigma'}c_{j\sigma}$$

$$+ P\sum_{<ij>}(c^{+}_{i\uparrow}c^{+}_{i\downarrow}c_{j\downarrow}c_{j\uparrow} + H.C.) + K\sum_{<ij>\sigma\sigma'} c^{+}_{i\sigma}c^{+}_{i\sigma'}c_{j\sigma'}c_{i\sigma} \quad (3)$$

Here symbols have their usual meaning. The first term represents the conventional hopping energy of electrons. Second term is on-site Coulomb interaction between opposite spin electrons $\left(U = \left\langle ii\left|\frac{1}{r}\right|ij\right\rangle\right)$. Third term is the Coulomb repulsion between neighboring electrons $\left( V = \left\langle ij \left| \frac{1}{r} \right| ji \right\rangle \right)$. Fourth term represents the exchange energy $\left( J = \left\langle ij \left| \frac{1}{r} \right| ij \right\rangle \right)$, while the fifth one is the pair hopping term $\left( P = \left\langle ii \left| \frac{1}{r} \right| jj \right\rangle \right)$. The last one is the hybrid interaction term $\left( K = \left\langle ii \left| \frac{1}{r} \right| ij \right\rangle \right)$.

The equation of motion of the Green's function

$$G_{ij}^{\sigma}(E) = \ll c_{i\sigma}; c_{j\sigma}^{+} \gg_E \tag{4}$$

is

$$E \ll c_{i\sigma}; c_{j\sigma}^{+} \gg = \frac{1}{2\pi} \left\langle [c_{i\sigma}, c_{j\sigma}^{+}] \right\rangle + \ll [c_{i\sigma}, H]; c_{j\sigma}^{+} \gg \tag{5}$$

which for the Hamiltonian H (Eq. (3)) becomes

$$\begin{aligned} E \ll c_{i\sigma}; c_{j\sigma}^{+} \gg = \frac{1}{2\pi}\delta_{ij} - t\sum_{l} \ll c_{l\sigma}; c_{j\sigma}^{+} \gg + U \ll n_{i-\sigma} c_{i\sigma}; c_{j\sigma}^{+} \gg \\ + J\sum_{l\sigma'} \ll c_{l\sigma'}^{+} c_{i\sigma'} c_{l\sigma}; c_{j\sigma'}^{+} \gg + V\sum_{l\sigma'} \ll c_{l\sigma'}^{+} c_{l\sigma'} c_{i\sigma}; c_{j\sigma'}^{+} \gg \\ + P\sum_{l\sigma'} \ll c_{i\sigma'}^{+} c_{l\sigma'} c_{l\sigma}; c_{j\sigma}^{+} \gg + K\sum_{l\sigma'} \ll c_{i\sigma'}^{+} c_{l\sigma'} c_{i\sigma}; c_{j\sigma}^{+} \gg \\ + K\sum_{l\sigma'} \ll c_{i\sigma'}^{+} c_{i\sigma'} c_{l\sigma}; c_{j\sigma}^{+} \gg \end{aligned} \tag{6}$$

Within mean-field approximation it gives

$$\begin{aligned} EG_{ij}^{\sigma}(E) = \frac{1}{2\pi}\delta_{ij} - t\sum_{l} G_{lj}^{\sigma}(E) + U < n_{i-\sigma} > G_{ij}^{\sigma} \\ + \sum_{l}{}' \left[ J\sum_{\sigma'} < c_{l\sigma'}^{+} c_{i\sigma'} > - V < c_{l\sigma}^{+} c_{i\sigma} > + P < c_{i-\sigma}^{+} c_{l-\sigma} > + K\sum_{\sigma'} < c_{i\sigma'}^{+} c_{i\sigma'} > \right] G_{lj}^{\sigma} \\ + \sum_{l}{}' \left[ -J < c_{l\sigma}^{+} c_{l\sigma} > + V\sum_{\sigma'} < c_{l\sigma'}^{+} c_{l\sigma'} > + K\sum_{\sigma'} < c_{i\sigma'}^{+} c_{l\sigma'} \right] G_{ij}^{\sigma} \\ + \sum_{l}{}' \left[ - < c_{i\sigma}^{+} c_{i\sigma} > G_{lj}^{\sigma} - < c_{i\sigma}^{+} c_{l\sigma} > \right] G_{ij}^{\sigma} \end{aligned} \tag{7}$$

Here, $\sum_{l}{}'$ means l sites are nearest neighbors to site i. Transforming Eq. (7) into k-space, one finds

$$EG_k^\sigma(E) = \frac{1}{2\pi} - t\sum_\delta e^{ik.\delta} G_k^\sigma(E) + Un_{-\sigma} G_k^\sigma(E)$$

$$+ \sum_\delta \left[ J\sum_{\sigma'} I_{\sigma'} - VI_\sigma + PI_{-\sigma} + Kn_{-\sigma} \right] e^{ik.\delta} G_k^\sigma$$

$$+ \sum_\delta \left[ -Jn_\sigma + V\sum_{\sigma'} n_{\sigma'} + KI_{-\sigma} \right] G_k^\sigma \qquad (8)$$

For nearest neighbor hopping the dispersion relation is $\varepsilon_k = -t\sum_\delta e^{+ik.\delta}$, so $\sum_\delta e^{ik\delta} = -\frac{\varepsilon_k}{t}$.

Also, let us use the notation $D = 2zt$, where z is the number of nearest neighbor. So Eq. (8) gives for the quasi-particle energy

$$E_\sigma(\varepsilon_k) = \left[ 1 - 2I_\sigma\left(\frac{zJ}{D} - \frac{zV}{D}\right) - 2I_{-\sigma}\left(\frac{zJ}{D} + \frac{zP}{D}\right) - 2n_{-\sigma}\frac{zK}{D} \right] \varepsilon_k$$

$$\left\{ -\sigma m \frac{U + zJ}{2} + zKI_{-\sigma} - \mu \right\} \qquad (9)$$

Constant term $\frac{U - zJ + 2zV}{2} n$ has been dropped.

With the notations similar to those of Hirsch [1], viz.

$$j_1 = \frac{zJ}{D} - \frac{zV}{D} \qquad (10a)$$

$$j_2 = \frac{zJ}{D} + \frac{zP}{D} \qquad (10b)$$

$$k_1 = \frac{U}{D} + \frac{zJ}{D} \qquad (10c)$$

$$k_2 = \frac{zK}{D} \qquad (10d)$$

Eq. (9) could be written as

$$E_\sigma(\varepsilon_k) = (1-2j_1 I_\sigma - 2j_2 I_{-\sigma} - 2k_2 n_{-\sigma})\varepsilon_k - \frac{\sigma D k_1}{2} M + D k_2 I_{-\sigma} - \mu \tag{11}$$

Here, $I_\sigma = < c^+_{i\sigma} c_{l\sigma} >$ (l is nearest neighbor to i)

Let $H_t$ denotes the hopping part of Hamiltonian

$$H_t = -t \sum_{<il>\sigma} c^+_{i\sigma}\, c_{l\sigma}$$

$$= -t \sum_{\delta} \sum_{k\sigma} e^{ik.\delta} n_{k\sigma}$$

$$= \sum_{k\sigma} \varepsilon_k n_{k\sigma} \qquad \left[ \text{as} - t\sum_{\delta} e^{ik.\delta} = \varepsilon_k \right]$$

$$\therefore \qquad -t \sum_{<il>\sigma} < c^+_{i\sigma} c_{l\sigma} > = \sum_{k\sigma} \varepsilon_k < n_{k\sigma} >$$

or $$I_\sigma = \frac{1}{(D/2)} \frac{1}{N} \sum_k (-\varepsilon_k) < n_{k\sigma} > \tag{12}$$

In terms of density of states $\rho(\varepsilon)$, $I_\sigma$ may be written as

$$I_\sigma = \int_{-D/2}^{D/2} \rho(\varepsilon) \left( -\frac{\varepsilon}{D/2} \right) d\varepsilon\, f(E_\sigma(\varepsilon)) \tag{13}$$

At T = 0 K

$$I_\sigma = \int_{-D/2}^{\epsilon_{F\sigma}} \rho(\varepsilon) \left( -\frac{\varepsilon}{D/2} \right) d\varepsilon \tag{14}$$

For square density of states

i.e., $\rho(\varepsilon) = \frac{1}{D}$ for $-D/2 \le \varepsilon \le D/2$

$$I_\sigma = -\frac{2}{D^2} \int_{-D/2}^{\varepsilon_{F\sigma}} \varepsilon\, d\varepsilon = \frac{2}{D^2} \left[ \frac{D^2}{8} - \frac{\varepsilon_{F\sigma}^2}{2} \right] \tag{15}$$

Now, $$n_\sigma = \int_{-D/2}^{\varepsilon_{F\sigma}} \rho(\varepsilon)\, d\varepsilon = \frac{\varepsilon_{F\sigma}}{D} + \frac{1}{2} \tag{16}$$

$$\begin{cases} n = n_\uparrow + n_\downarrow & \text{(17a)} \\ M = n_\uparrow - n_\downarrow & \text{(17b)} \end{cases}$$

From Eqs. (15) and (16) one finds

$$I_\sigma = n_\sigma(1 - n_\sigma) \tag{18}$$

## III. Phase Diagrams at T = 0 K

Let us consider, firstly, the case at T = 0K. In case the system is not completely polarized (i.e., the case of $M \neq n$), the chemical potential is governed by the condition

$$E_\uparrow(\varepsilon_{F\uparrow}) = E_\downarrow(\varepsilon_{F\downarrow}) \tag{19}$$

Using Eqs. (11), (16) and (18), condition (19) may be expressed as

$$(j_1 - j_2)(1-n)^2 - (j_1 + j_2)\frac{1 - M^2 - (1-n)^2}{2} = (k_1 - 1) - (n-2)k_2 \tag{20}$$

The coefficient of $\varepsilon_k$ in Eq. (9) decides the band narrowing (or broadening) due to interactions. In fact, this coefficient can not be negative, i.e.,

$$(1 - 2j_1 I_\sigma - 2j_2 I_{-\sigma} - 2k_2 n_{-\sigma}) > 0 \tag{21}$$

In Figs. 1(a) and 1(b), we show magnetic-nonmagnetic phase diagram as a function of the parameter $k_2$ and band filling n. Here, F denotes fully polarized ferromagnetic phase (M = n), P denotes paramagnetic phase (M = 0) and PF denotes partially ferromagnetic phase (0 < M < n). It may be noticed that curves in these figures are not symmetric around n = 1. In fact inter-site hybrid interactions break electron-hole symmetry (at least in mean-field approximation).

Figure 2 show lines of constant magnetization in $j_2 - k_2$ plane for quarter filled (n = 0.5) and half-filled (n = 1.0) bands, respectively in (a) and (b). Parameters $j_2$ and $k_2$ are supplementing each other for the on-set of ferromagnetism. The system is fully spin polarized above the line labeled M = 0.5 (for n = 0.5, Fig. (a)) and M = 1.0 (for n = 1.0, Fig. (b)). Below line M = 0, the system is paramagnetic. Figures 3(a) and 3(b) show behavior of saturation magnetization $M_s$ as a function of band filling n for various values of interaction parameters $k_2$ and $j_2$. It is clear from Fig. 3(a) ($j_2 = 1.8$) that magnetization is least favored around half band filling. In fact for $k_2 = 0.1$, the magnetization is zero around n = 1.0. For $k_2 = 0.2$ and 0.3, the magnetization M is much less than n around half filling. For $j_2 = 0.2$, we observe quite surprising results. For $k_2 = 0.6$, the magnetization M = n upto $n \cong 0.55$ and drops rapidly to zero for n > 0.55. For higher values of $k_2$ this drop (of magnetization from its saturation value to zero) takes place at higher values of band filling n. For $k_2 = 0.9$, the drop takes place around n = 1.0. Surprisingly similar behavior of magnetization as a function of d-band filling has been observed experimentally in the itinerant ferromagnetic systems $Fe_{1-x}Co_xS_2$ and $Co_{1-x}Ni_xS_2$ [10]. Also similar behavior of magnetization has been seen in the theoretical work of Didukh et al.[11] in their study of doubly orbitally degenerate generalized Hubbard model.

## IV. Study at Finite Temperature

At finite temperature, the magnetization m is obtained by solving

$$M = (n_\uparrow - n_\downarrow) = \int_{-D/2}^{D/2} d\varepsilon\, \rho(\varepsilon) \left\{ f(E_\uparrow(\varepsilon)) - f(E_\downarrow(\varepsilon)) \right\} \tag{22}$$

Self-consistently alongwith Eqs. (11) and (13). The variation of magnetization as a function of temperature is shown in Fig. 4 for various values of $k_2$ and $j_2$ for a quarter filled band.

The effective mass $M_\sigma^*$ for spin $\sigma$, could be expressed as inverse of effective bandwidth $D_\sigma^*$ for spin $\sigma$, from Eq. (11) as

$$\frac{m_\sigma^*}{m} = \frac{D}{D_\sigma^*} = (1 - 2j_1 I_\sigma - 2j_2 I_{-\sigma} - 2k_2 n_{-\sigma})^{-1} \tag{23}$$

Here $m$ and $D$ are the spin-independent bare mass and bandwidth. Figures. 5(a) and 5(b) show behavior of effective mass for majority (up) and minority (down) spin electrons as a function of temperature.

Using Eqs. (13) and (22) in presence of small constant external magnetic field and expanding Fermi functions $f(E_\sigma(\varepsilon))$ around chemical potential one finds for $(I_\uparrow - I_\downarrow)$ and m following expressions

$$I_\uparrow - I_\downarrow = \left\{(j_2 - j_1)(I_\uparrow - I_\downarrow) + k_2 M\right\} G_2 + \left\{k_1 M + 2h + (I_\uparrow - I_\downarrow) k_2\right\} G_1 \tag{24a}$$

$$M = \left\{(j_2 - j_1)(I_\uparrow - I_\downarrow) + k_2 M\right\} G_1 + \left\{k_1 M + 2h + (I_\uparrow - I_\downarrow) k_2\right\} G_o \tag{24b}$$

where

$$G_l = D \int_{-D/2}^{D/2} d\varepsilon\, \rho(\varepsilon) \left(-\frac{\varepsilon}{D/2}\right)^l \left(-\frac{\partial f}{\partial \varepsilon_\sigma}\right) \tag{24c}$$

In the paramagnetic phase, the magnetic susceptibility $\chi$ is obtained from (24b)

$$\chi = \left.\frac{\partial M}{\partial h}\right|_{n=0} = \frac{2G_1^2(j_2 - j_1) + 2k_2 G_o G_1 A + 2G_o[1 - (j_2 - j_1)G_2 - k_2 G_1]}{A\,[1 - (j_2 - j_1)G_2 - k_2 G_1] - (j_2 - j_1)G_1(k_2 G_2 + kG_1) - k_2 G_o (k_2 G_2 + kG_1)\, A} \tag{24d}$$

where

$$A = 1 - k_1 G_o - k_2 G_1 \tag{24e}$$

Figures 6(a) and 6(b) show variation of inverse susceptibility as a function of temperature.

The critical temperature $T_c$ may be calculated from the following equation obtained by setting to use the determinant of the coefficients of $(I_\uparrow - I_\downarrow)$ and m in equations (24a) and (24b).

$$\left\{1 - (j_2 - j_1)G_2 - k_2 G_1\right\}\left\{1 - k_1 G_o - k_2 G_1\right\} - \left\{k_1 G_1 + k_2 G_2\right\}\left\{(j_2 - j_1)G_1 + k_2 G_o\right\} = 0 \tag{25}$$

Effective magnetic moment $p_{eff}$ in the itinerant electron system may be obtained from the magnetic susceptibility $\chi$ (Eq. (24d)), if we express it in the form of susceptibility of a localized moment system as T approaches $T_c$; viz.

$$\chi = \frac{p_{eff}\ (T)}{3\,(T - T_c)} \tag{26}$$

Variation of effective magnetic moment with temperature is shown in Figs. 7(a) and 7(b). The real part of a.c. (optical) conductivity is given as

$$\sigma_1(\omega) = \frac{ne^2}{m*}\frac{\tau}{1 + \omega^2\tau^2} \tag{27}$$

Figure 8 shows variation of optical conductivity $\sigma_1(\omega)$ as a function of frequency. Assuming that there is no significant change in the relaxation time with spin polarization, the optical weight W (T, H) may be obtained from

$$W(T, H) = \frac{2}{\pi e^2}\int_0^{\omega_m} \sigma_1(\omega)\, d\omega = \frac{n}{m*} \tag{28}$$

Figure 9(a) shows behavior of optical weights $\frac{n_\sigma}{m_\sigma^*}$ for spin $\sigma$ electrons as a function of temperature. Figure 9(b) shows optical weights for up and down spins combined. Singley et al. [12] have recently performed infrared spectroscopic measurements on itinerant ferromagnet $Ga_{1-x}Mn_xAs$. Their results of spectral weights, for ferromagnetic sample, resemble qualitatively with other results of total spectral weight for h = 0 (i.e., in absence of applied magnetic field).

Under the assumption of a constant relaxation time, the magnetoresistance may be estimated from the change in effective mass with spin polarization

$$\frac{\Delta\rho}{\rho} = \frac{\rho(H) - \rho(0)}{\rho(0)} =$$

$$\left[\frac{\sum_{\sigma} n_{\sigma}(H)\{1-2j_1 I_{\sigma}(H)-2j_2 I_{-\sigma}(H)-2k_2 n_{-\sigma}(H)\}}{\sum_{\sigma} n_{\sigma}(0)\{1-2j_1 I_{\sigma}(0)-2j_2 I_{-\sigma}(0)-2k_2 n_{-\sigma}(0)\}}-1\right] \tag{29}$$

Figure 10 shows the behavior of magnetoresistance as a function of temperature for different magnetic fields.

## V. Discussions

In this work we have revisited a model for the origin of ferromagnetism where alongwith exchange splitting (as a result of intra-site Coulomb interaction U), the intersite exchange and hybrid interactions play a crucial role. In the recent past, the role of hybrid interactions has been studied by many workers [1,5,13-16]. We have studied various properties like magnetic-nonmagnetic phase diagram, saturation magnetization versus band filling at T = 0 K. Also studied are the variation of magnetic moment (below $T_c$), magnetic susceptibility (above $T_c$), electronic effective mass, as a function of temperature. And finally we have studied optical conductivity $\sigma_1(\omega)$, optical weights $\frac{W(T, H)}{n/m^*}$ and magnetoresistance $\rho(H)$ at various temperatures.

We have studied explicitly the effect of intersite hybrid interactions $k_2$ $\left(k_2 = \frac{zK}{D}, K = \left\langle ii \left| \frac{1}{r} \right| ij \right\rangle\right)$ in controlling all above mentioned physical quantities. It is clear from Figs. (1) – (3) that parameter $k_2$ (hybrid interactions) play a dominant role in producing ferromagnetism. Fig. 1(b) shows that F phase (the fully polarized ferromagnetic phase) can be present (for larger values of $k_2$) even for very small values of $j_2$.

Role of parameter $k_2$ is explicitly seen in all the physical quantities studied in this paper. And with the fact that systems under consideration (e.g. transition metals) the hybrid interactions $k_2$ have larger magnitude than the exchange interactions $j_2$, this study of the role played by $k_2$ is all the more important.

**Acknowledgement:** P.D. and I.S. are thankful to the Department of Science and Technology, N. Delhi (India) for financial support.

## List of Figure Captions

**Figure 1** Phase diagram at T = 0K in $k_2$ – n plane. P, PF and F denote respectively paramagnetic, partially ferromagnetic and fully ferromagnetic phases.

**(a)** for $j_2$ = 1.8, **(b)** for $j_2$ = 0.2.

**Figure 2** Lines of Constant magnetization in $k_2$ – $j_2$ plane.

**(a)** for n = 0.5, **(b)** for n = 1.0.

**Figure 3** Saturation Magnetization as a function of band filling n.

**(a)** for $j_2$ = 1.8, **(b)** $j_2$ = 0.2.

**Figure 4** Magnetization versus temperature for quarter filled band (n = 0.5).

**(a)** for $j_2$ = 1.8, **(b)** $j_2$ = 0.2

**Figure 5** Effective mass ratio versus temperature. Dotted lines for up spin and full lines for down spin.

**(a)** for $j_2$ = 1.8, **(b)** for $j_2$ = 0.2.

**Figure 6** Inverse susceptibility versus tempeature.

**(a)** for $j_2$ = 1.8, $k_2$ = 0.1 **(b)** for $j_2$ = 0.2, $k_2$ = 0.8.

**Figure 7** Effective moment versus temperature.

**(a)** for $j_2$ = 1.8, $k_2$ = 0.1 **(b)** for $j_2$ = 0.2, $k_2$ = 0.8.

**Figure 8** Optical conductivity versus frequency (full lines in absence of magnetic field h and dotted lines for h = 0.05 for $j_2$ = 0.2, $k_2$ = 0.8.

**Figure 9 (a)** Optical weights for spin up (full lines) and spin down (dotted lines) for $j_2$ = 0.2, $k_2$ = 0.8.

**Figure 9 (b)** Optical weights for up and down spin combined for $j_2$ = 0.2, $k_2$ = 0.8.

**Figure 10** Magnetoresistance versus temperature for $j_2$ = 0.2, $k_2$ = 0.8.

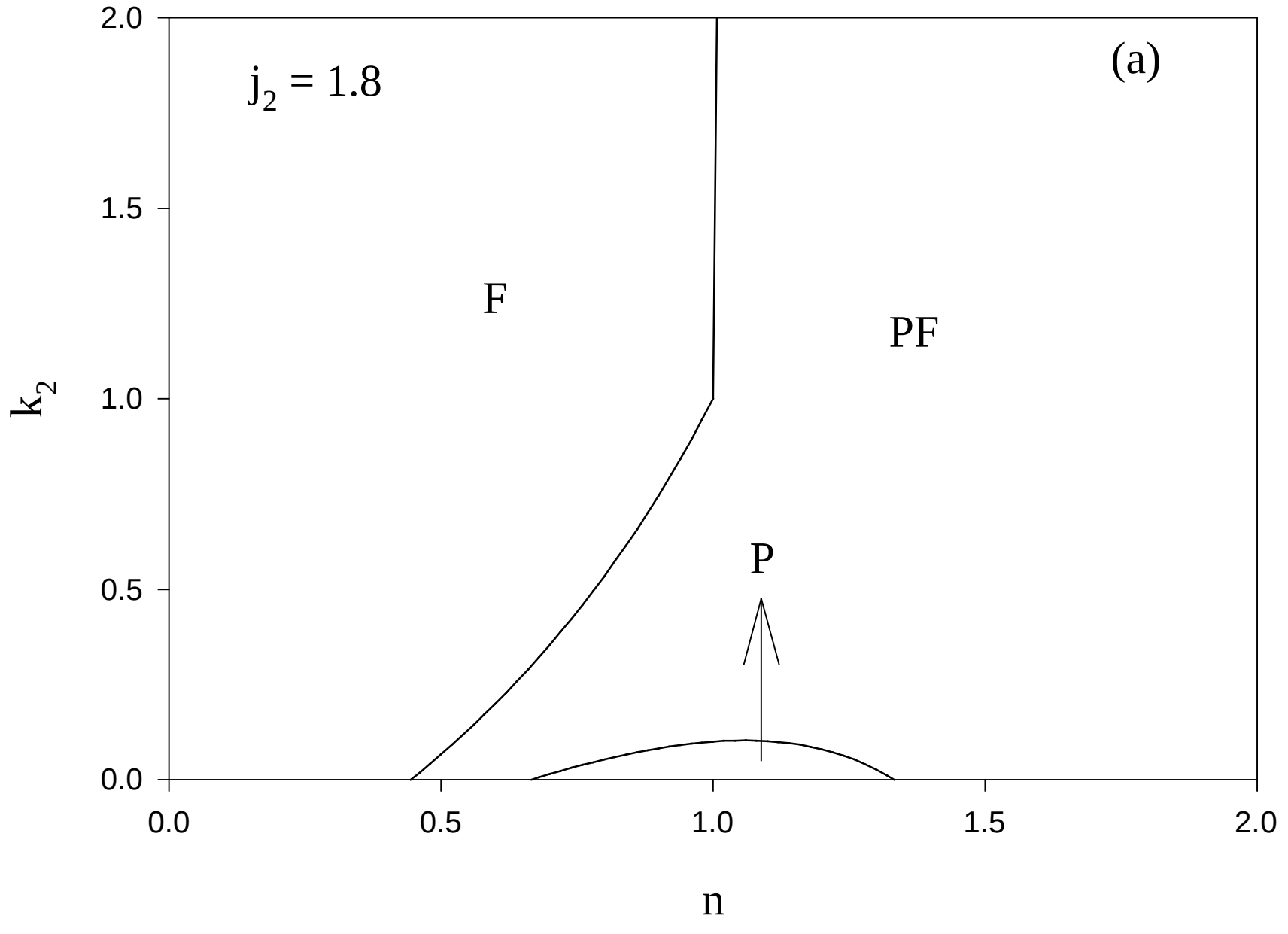

**Figure 1(a)**

Role of Inter-site Hybrid Interactions in Itinerant Ferromagnetism

Piyush Dua, Sunil Panwar and Ishwar Singh

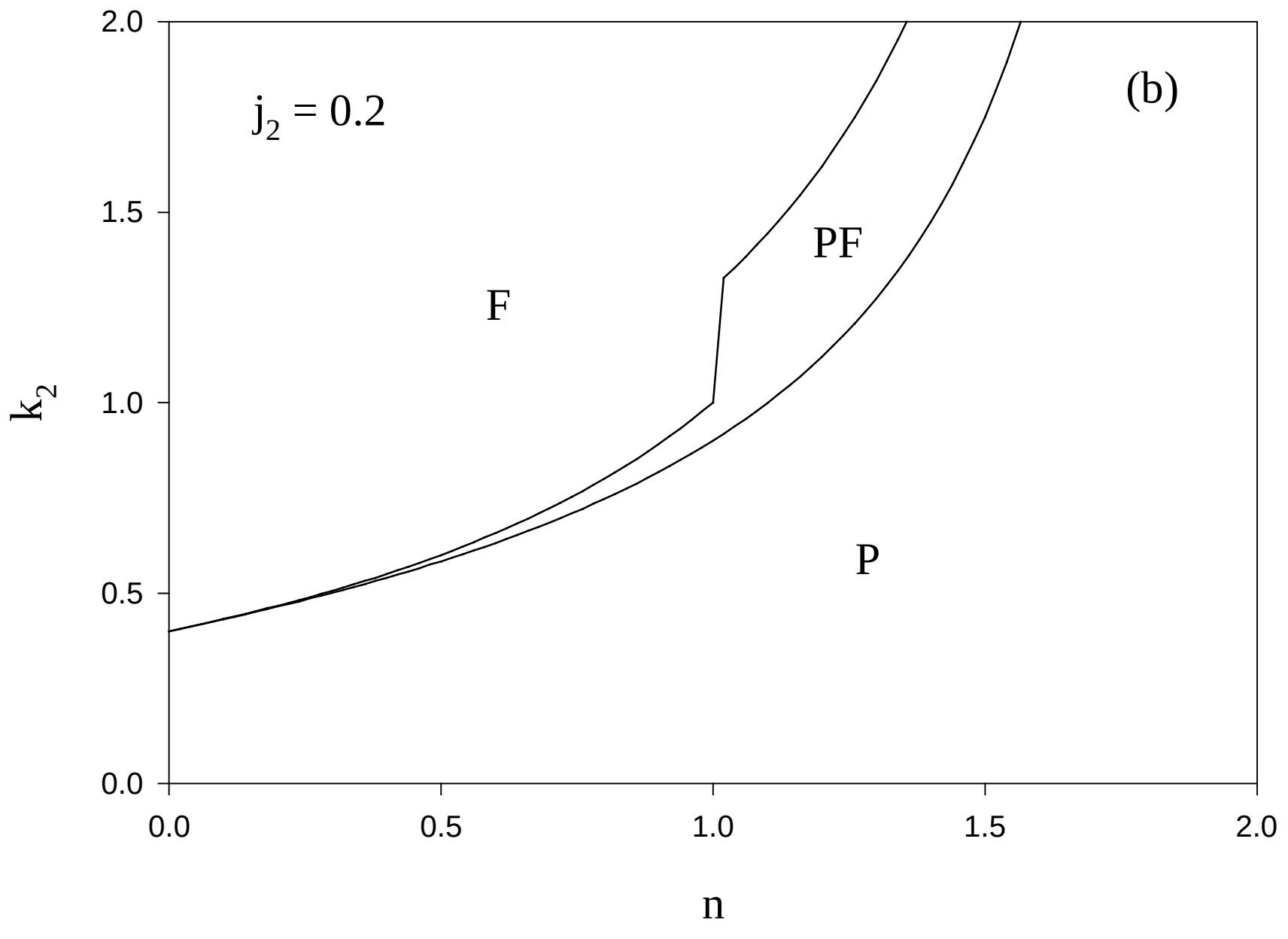

**Figure 1(b)**

Role of Inter-site Hybrid Interactions in Itinerant Ferromagnetism

Piyush Dua, Sunil Panwar and Ishwar Singh

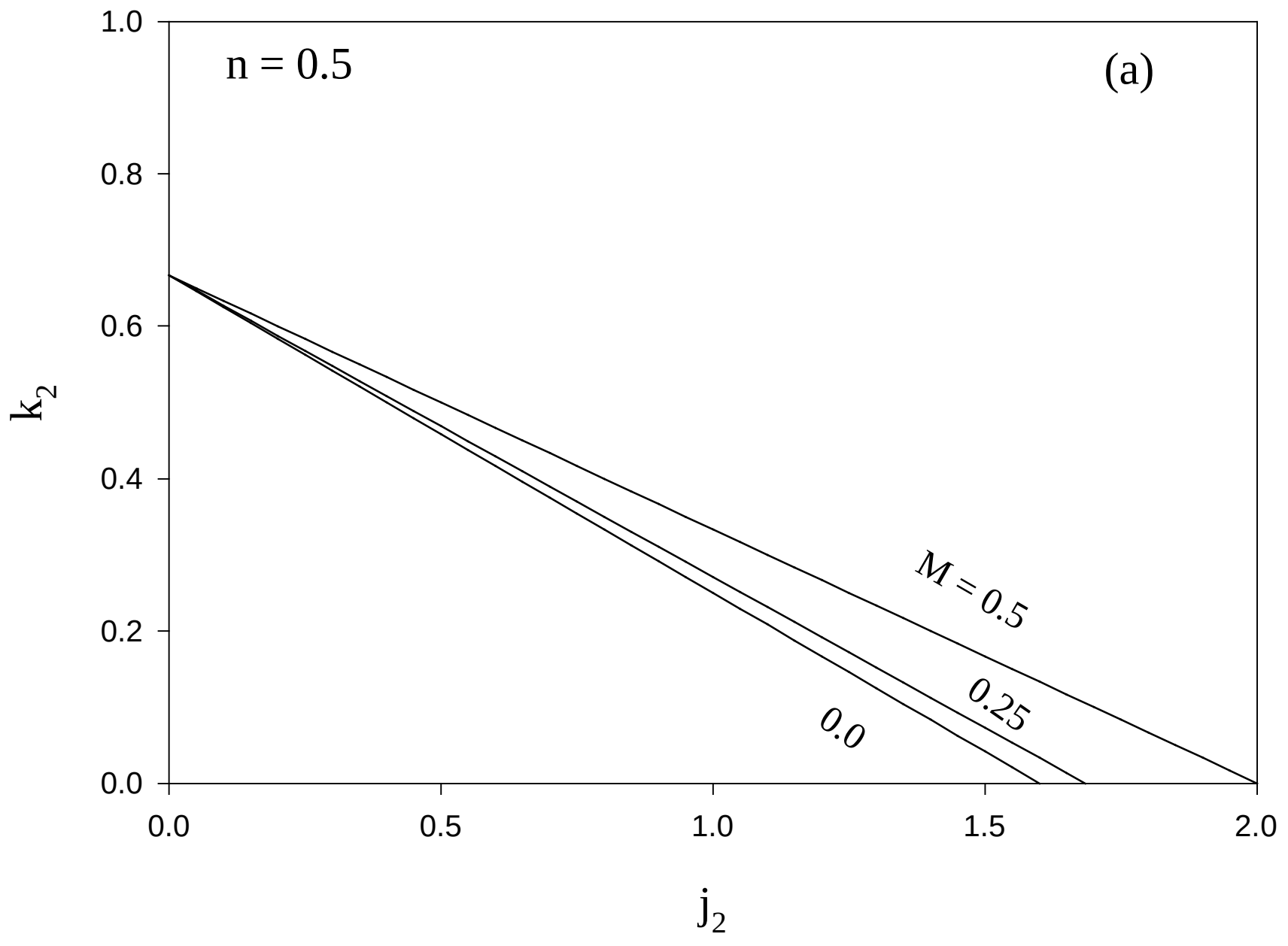

**Figure 2(a)**

Role of Inter-site Hybrid Interactions in Itinerant Ferromagnetism

Piyush Dua, Sunil Panwar and Ishwar Singh

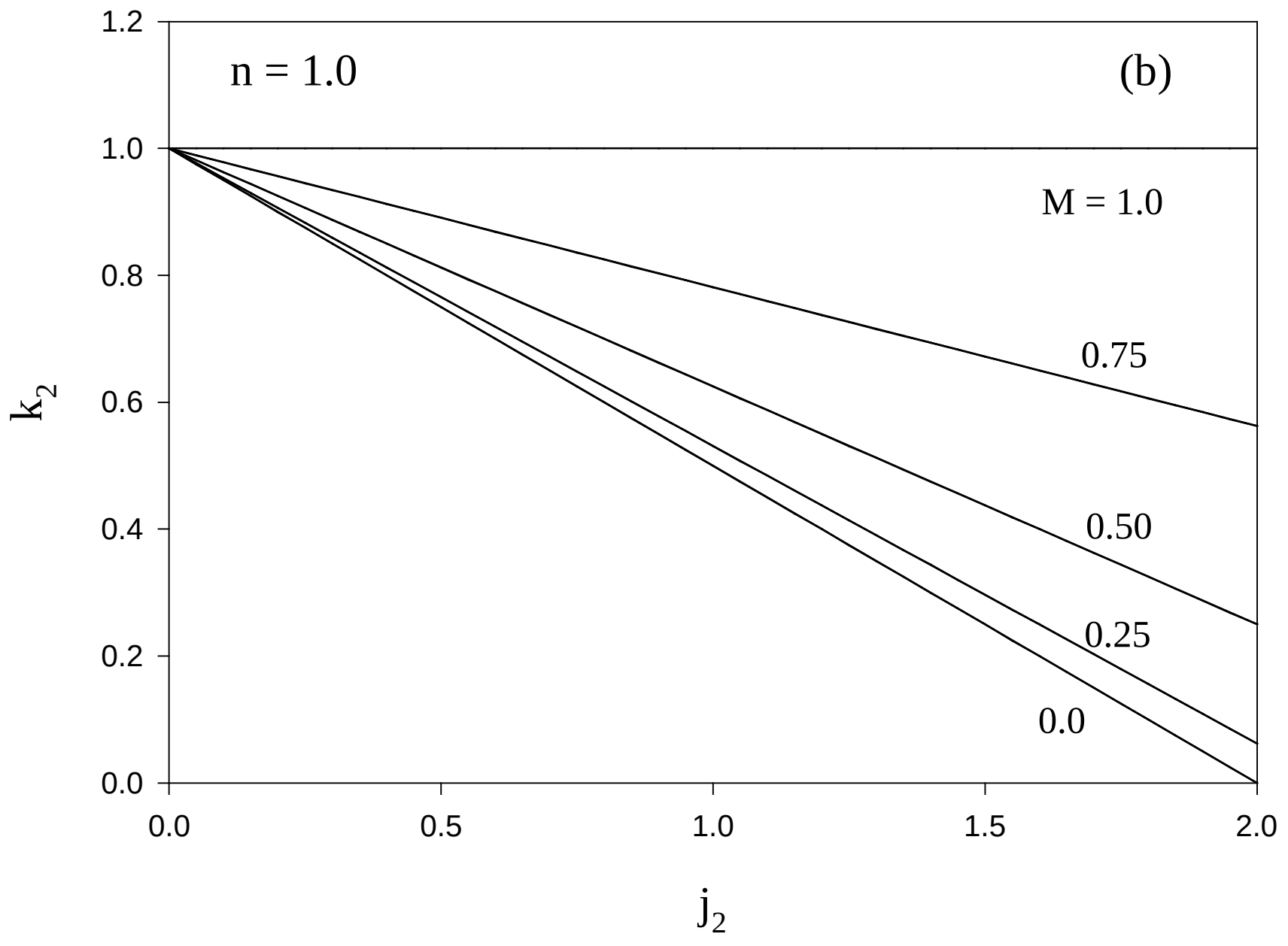

**Figure 2(b)**

Role of Inter-site Hybrid Interactions in Itinerant Ferromagnetism

Piyush Dua, Sunil Panwar and Ishwar Singh

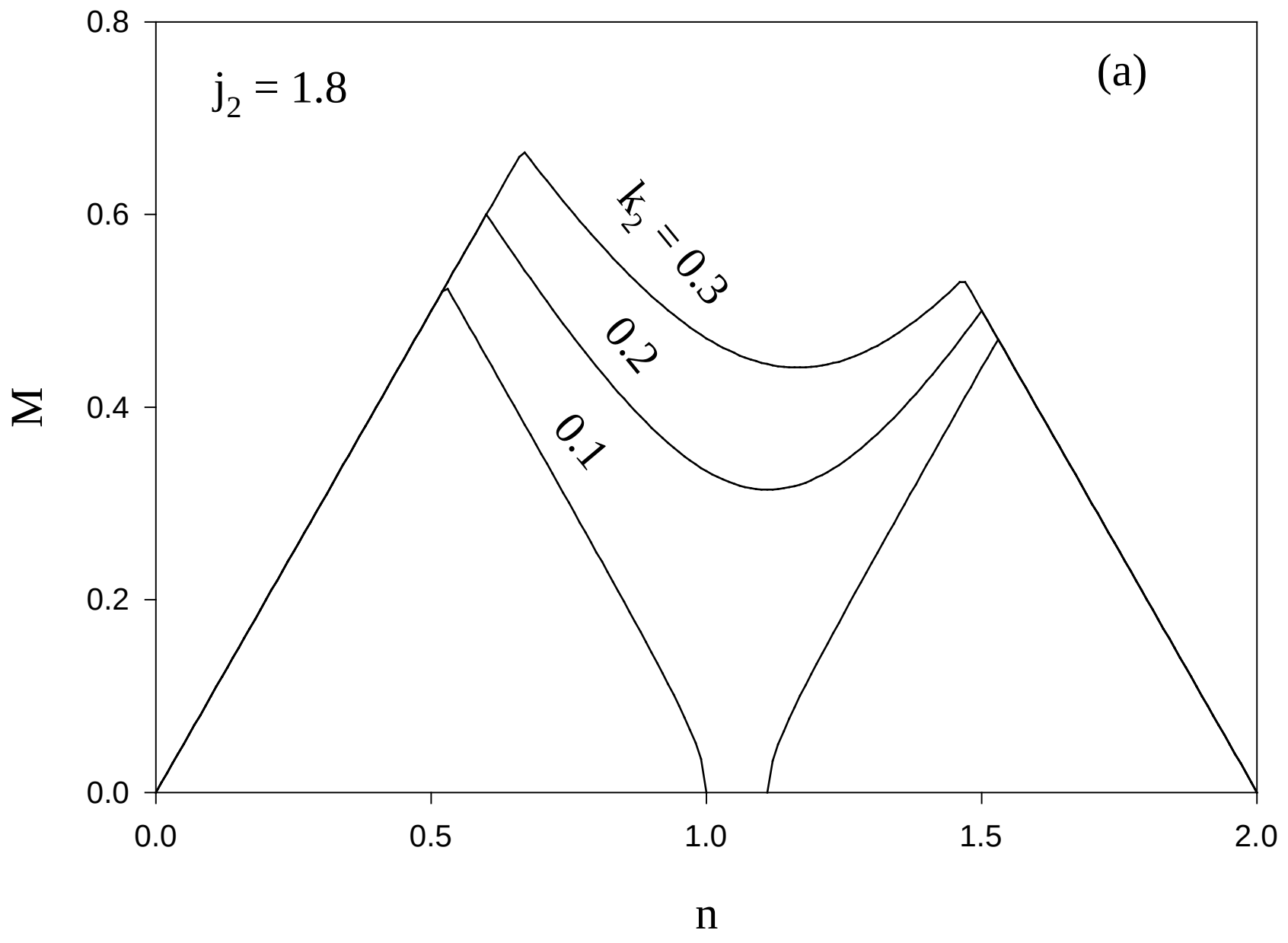

**Figure 3(a)**

Role of Inter-site Hybrid Interactions in Itinerant Ferromagnetism

Piyush Dua, Sunil Panwar and Ishwar Singh

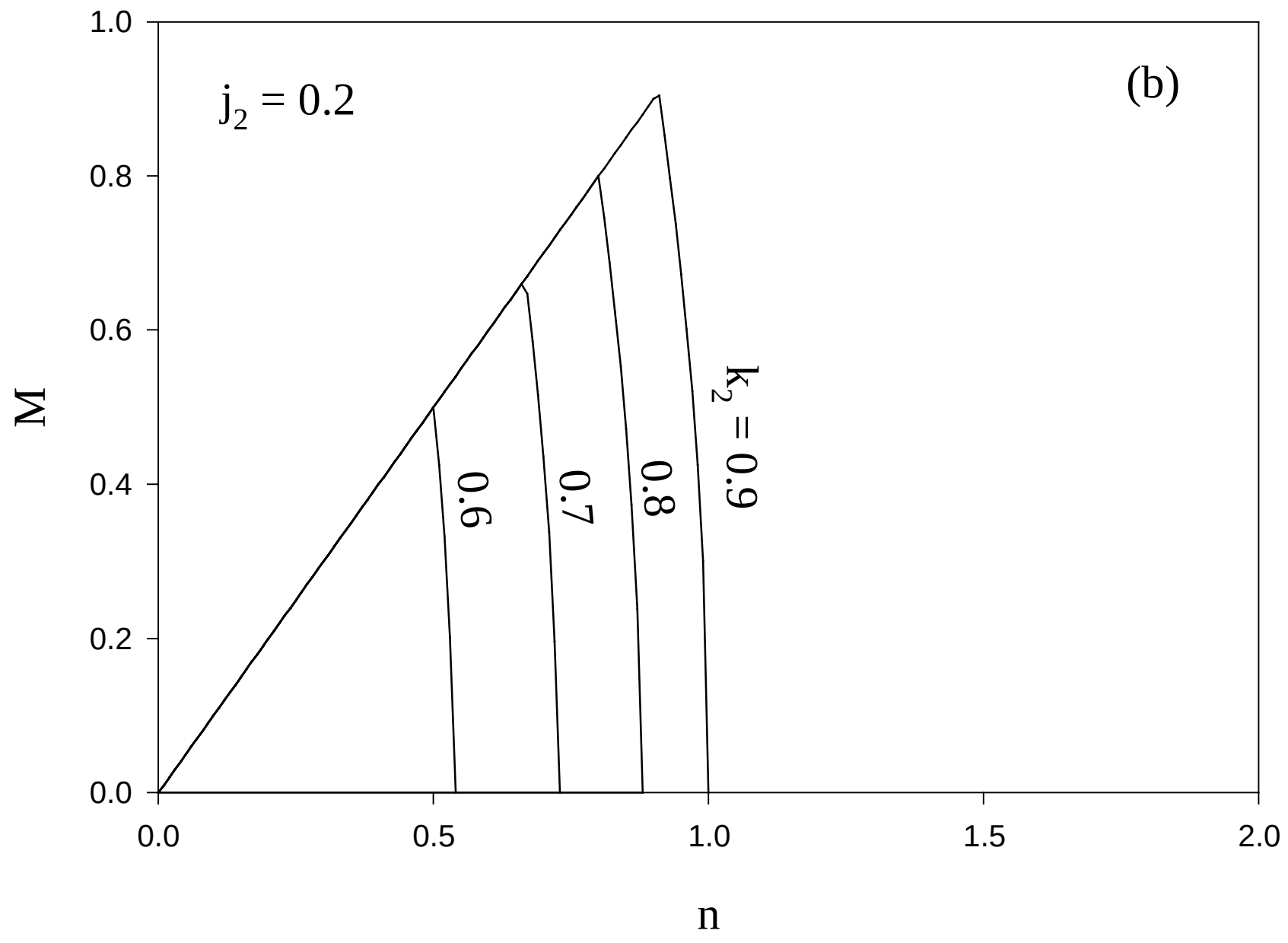

**Figure 3(b)**

Role of Inter-site Hybrid Interactions in Itinerant Ferromagnetism

Piyush Dua, Sunil Panwar and Ishwar Singh

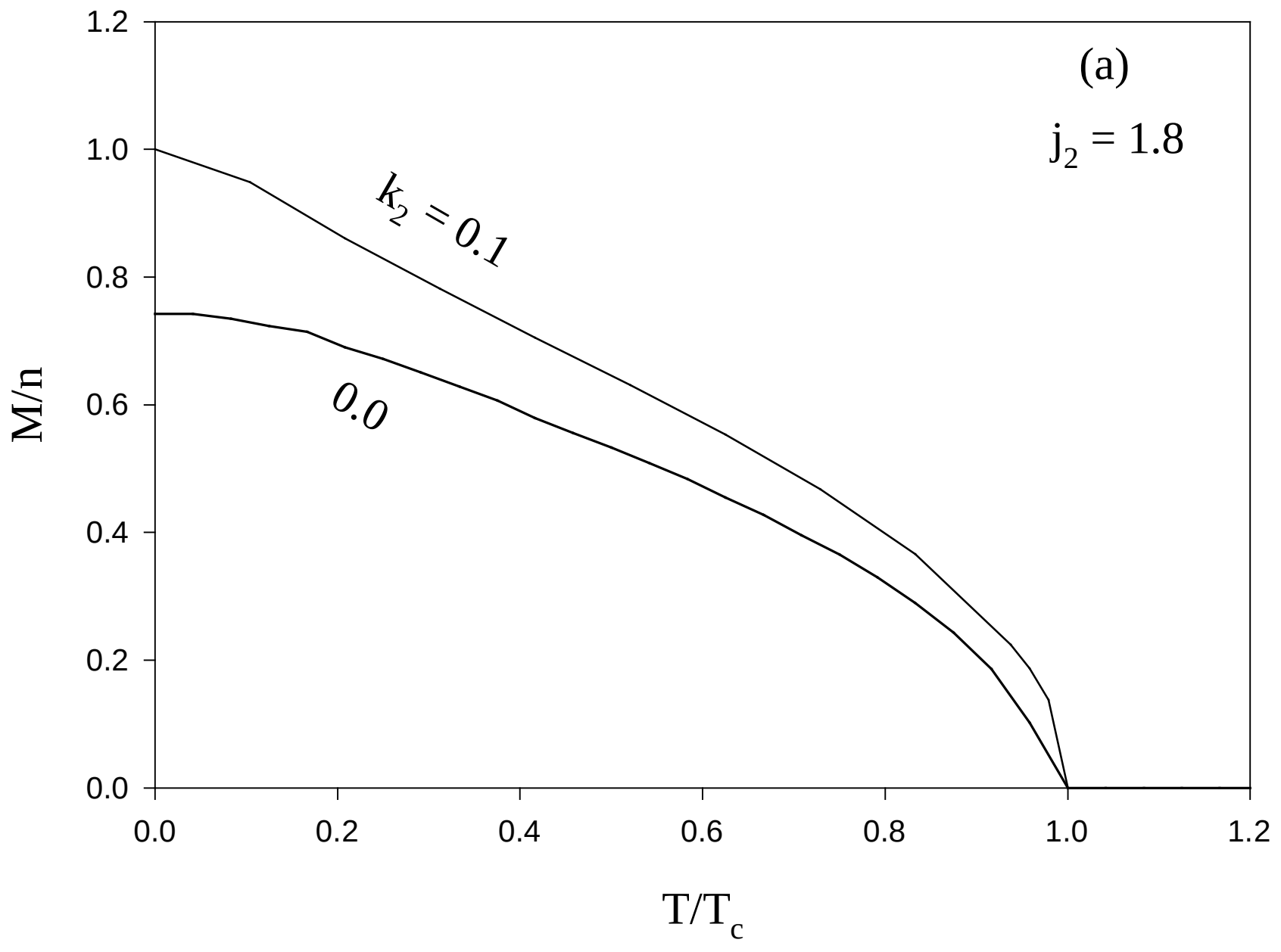

**Figure 4(a)**

Role of Inter-site Hybrid Interactions in Itinerant Ferromagnetism

Piyush Dua, Sunil Panwar and Ishwar Singh

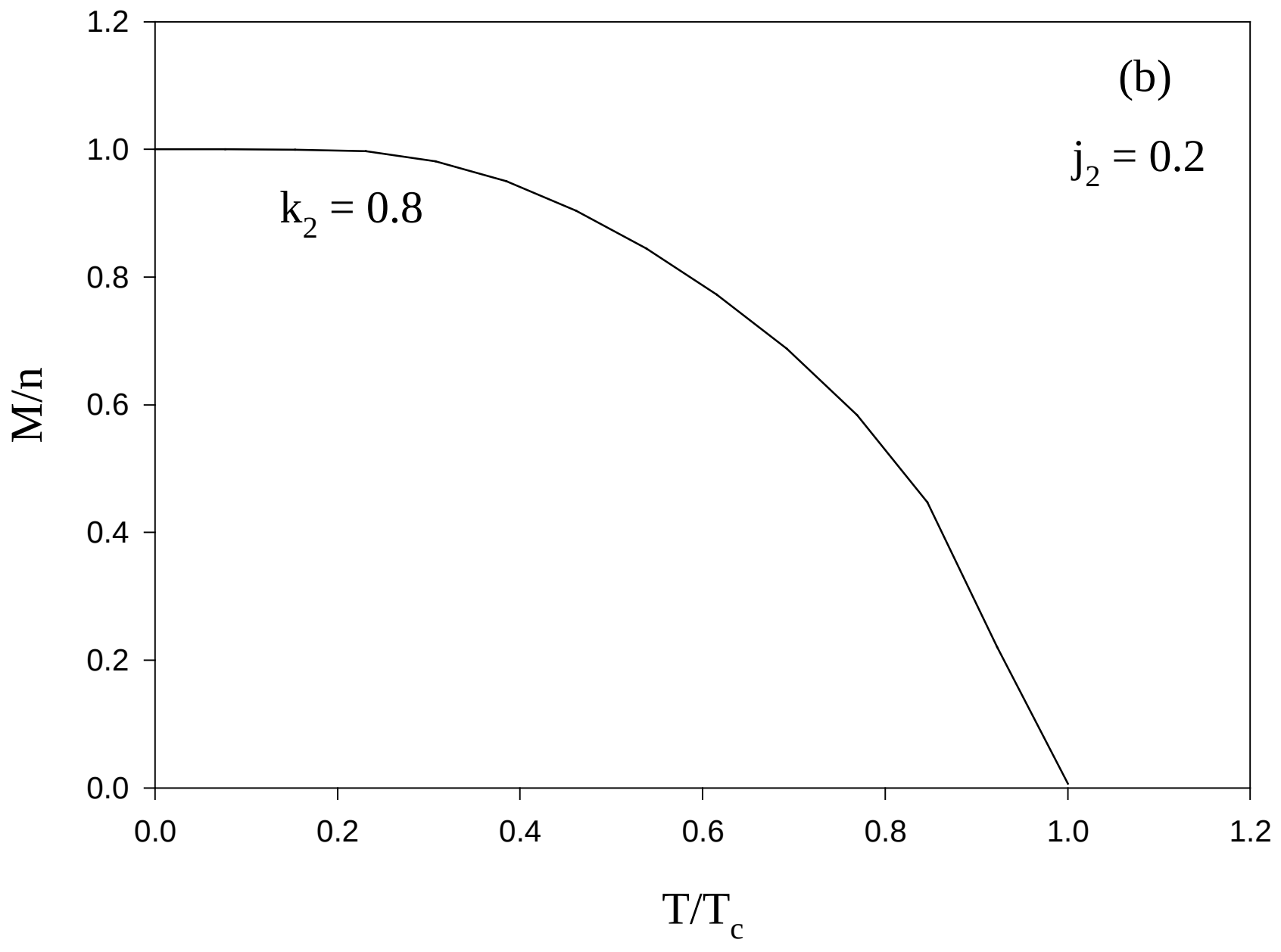

**Figure 4(b)**

Role of Inter-site Hybrid Interactions in Itinerant Ferromagnetism

Piyush Dua, Sunil Panwar and Ishwar Singh

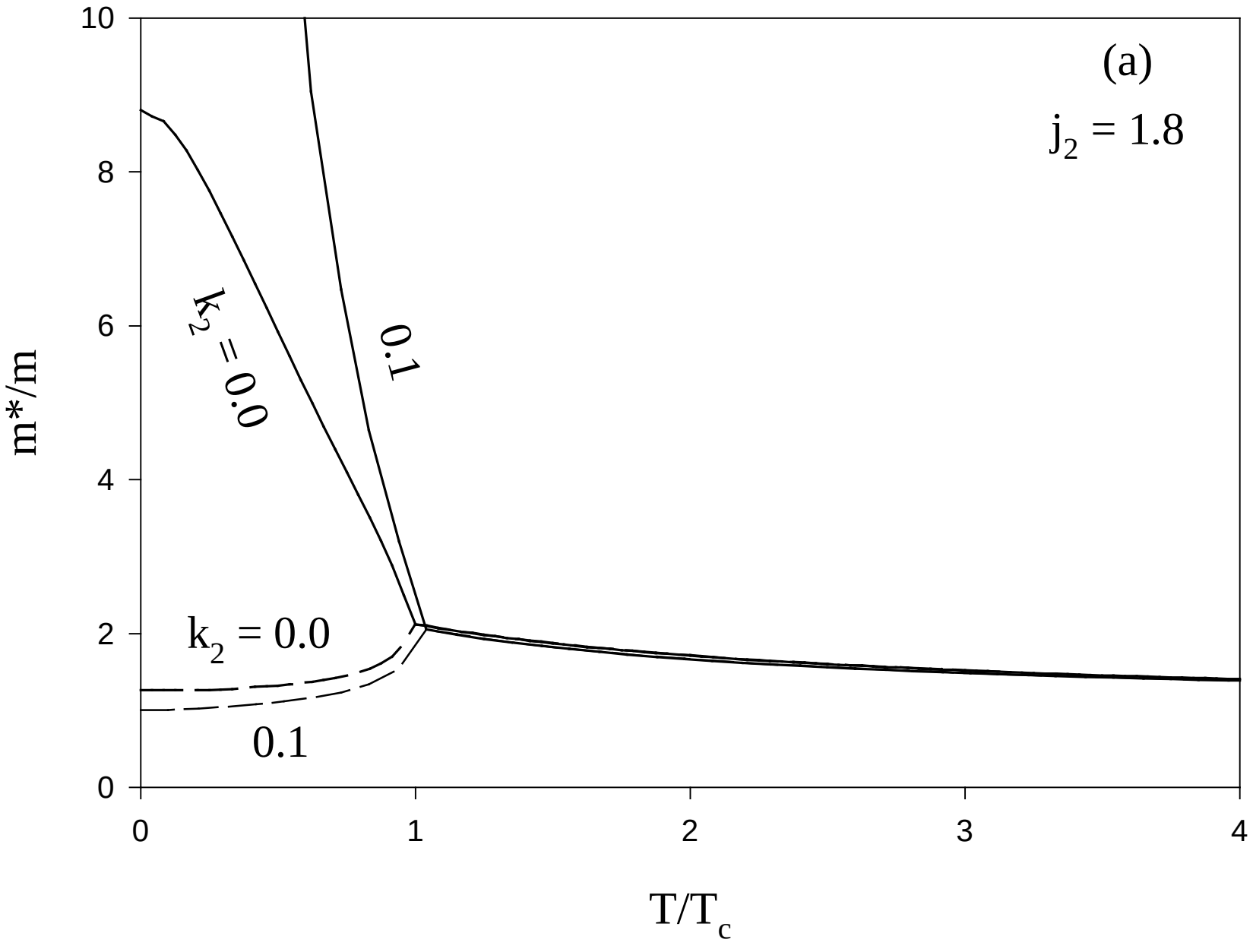

**Figure 5(a)**

Role of Inter-site Hybrid Interactions in Itinerant Ferromagnetism

Piyush Dua, Sunil Panwar and Ishwar Singh

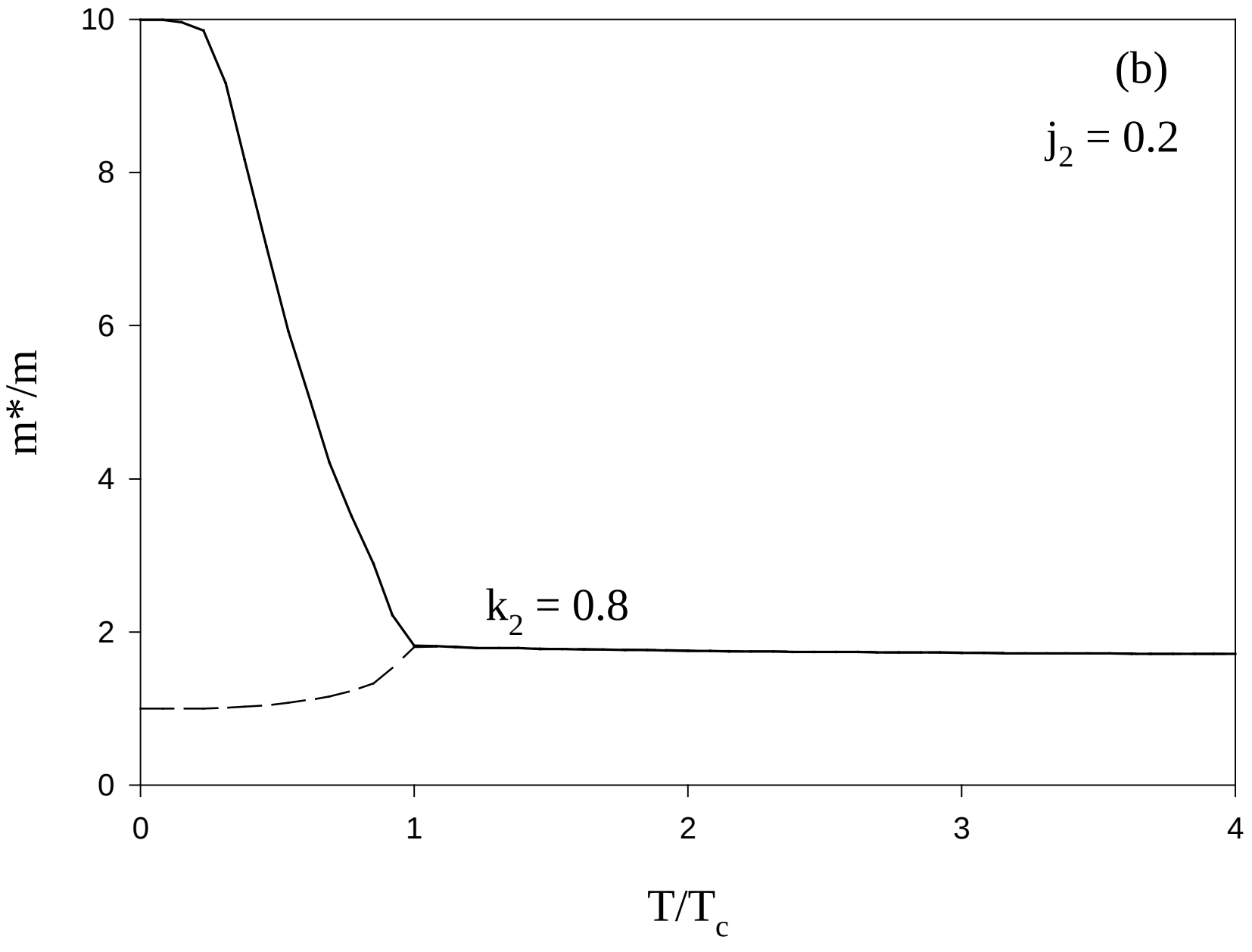

**Figure 5(b)**

Role of Inter-site Hybrid Interactions in Itinerant Ferromagnetism

Piyush Dua, Sunil Panwar and Ishwar Singh

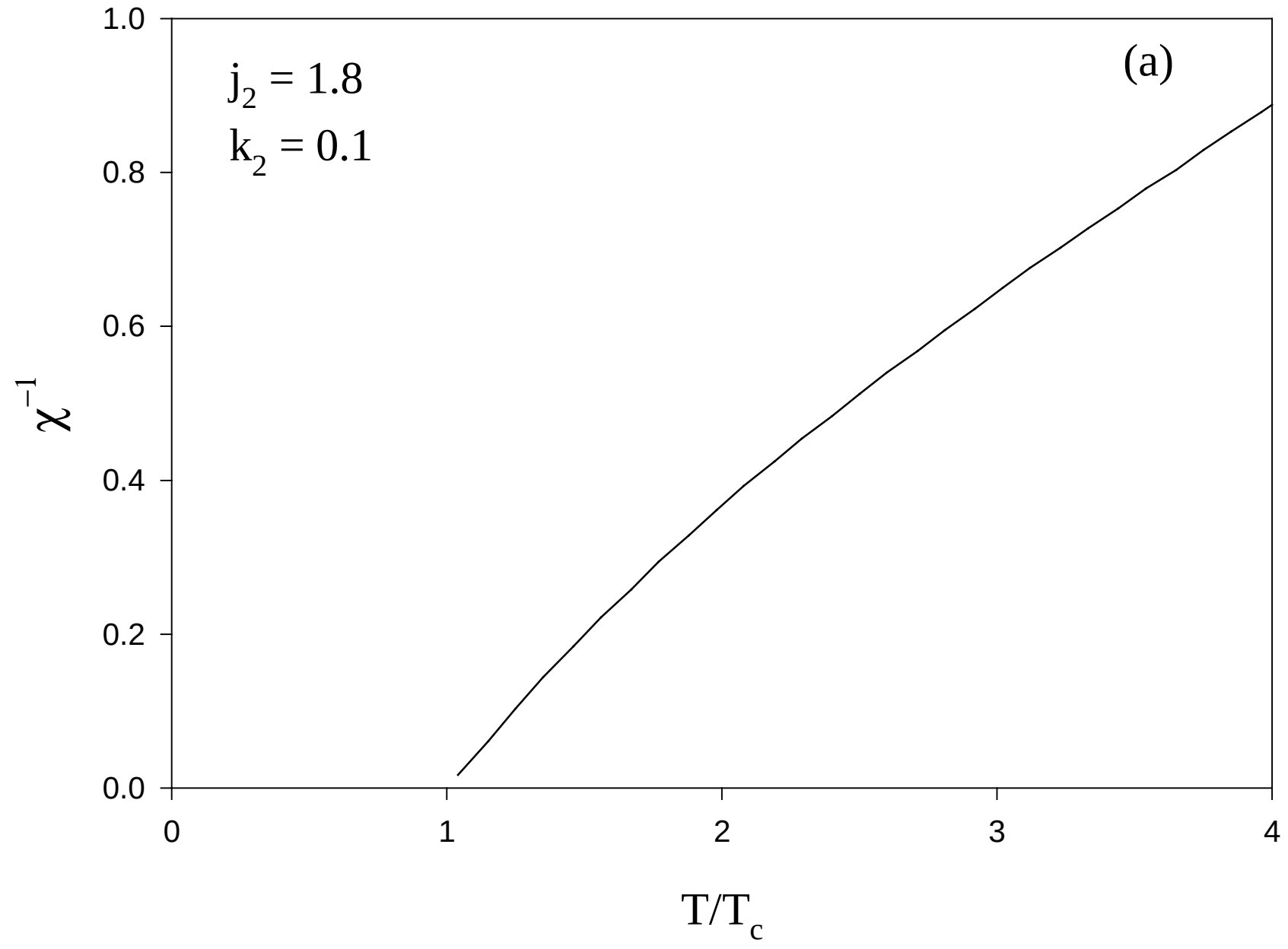

**Figure 6(a)**

Role of Inter-site Hybrid Interactions in Itinerant Ferromagnetism

Piyush Dua, Sunil Panwar and Ishwar Singh

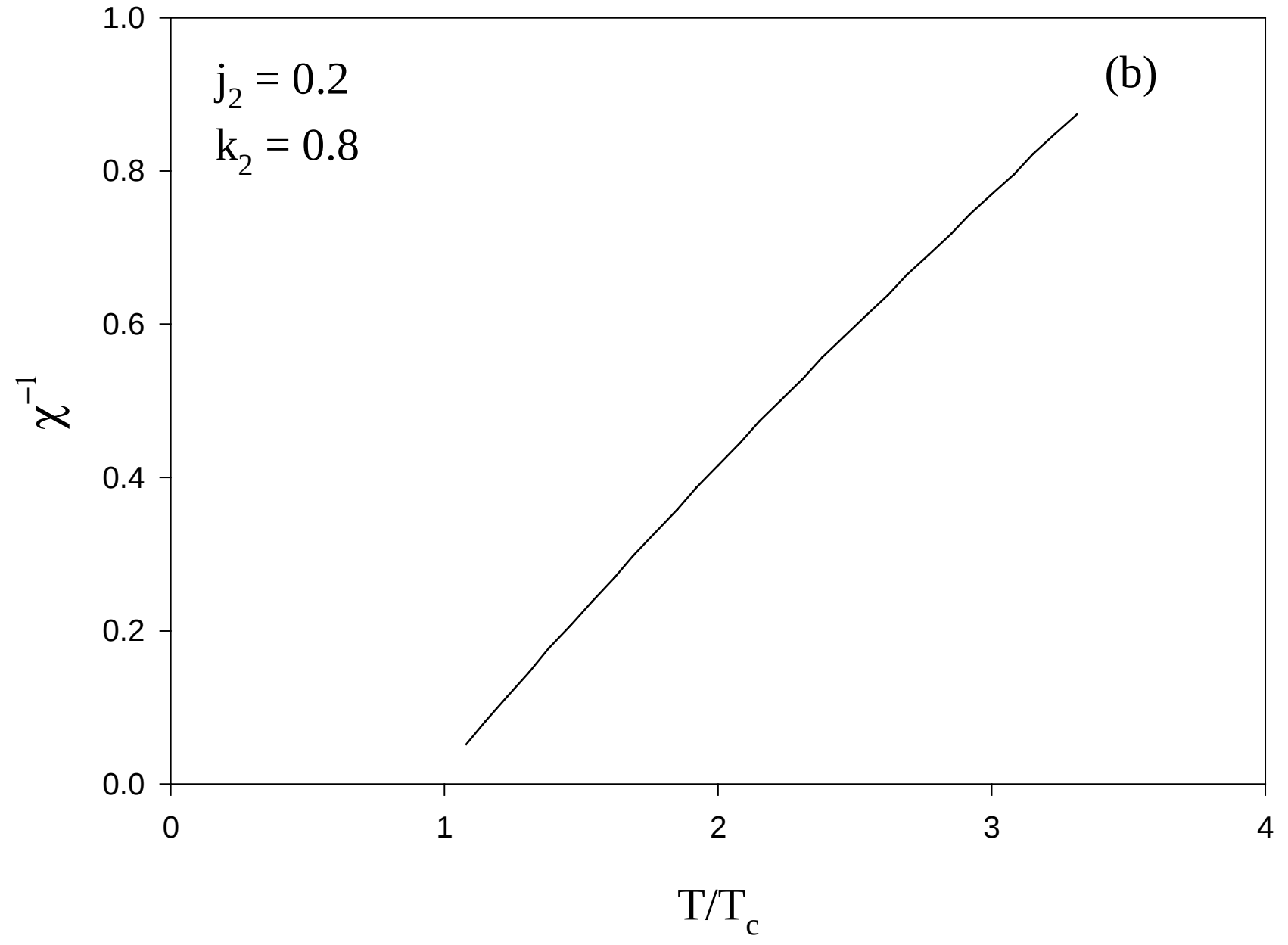

**Figure 6(b)**

Role of Inter-site Hybrid Interactions in Itinerant Ferromagnetism

Piyush Dua, Sunil Panwar and Ishwar Singh

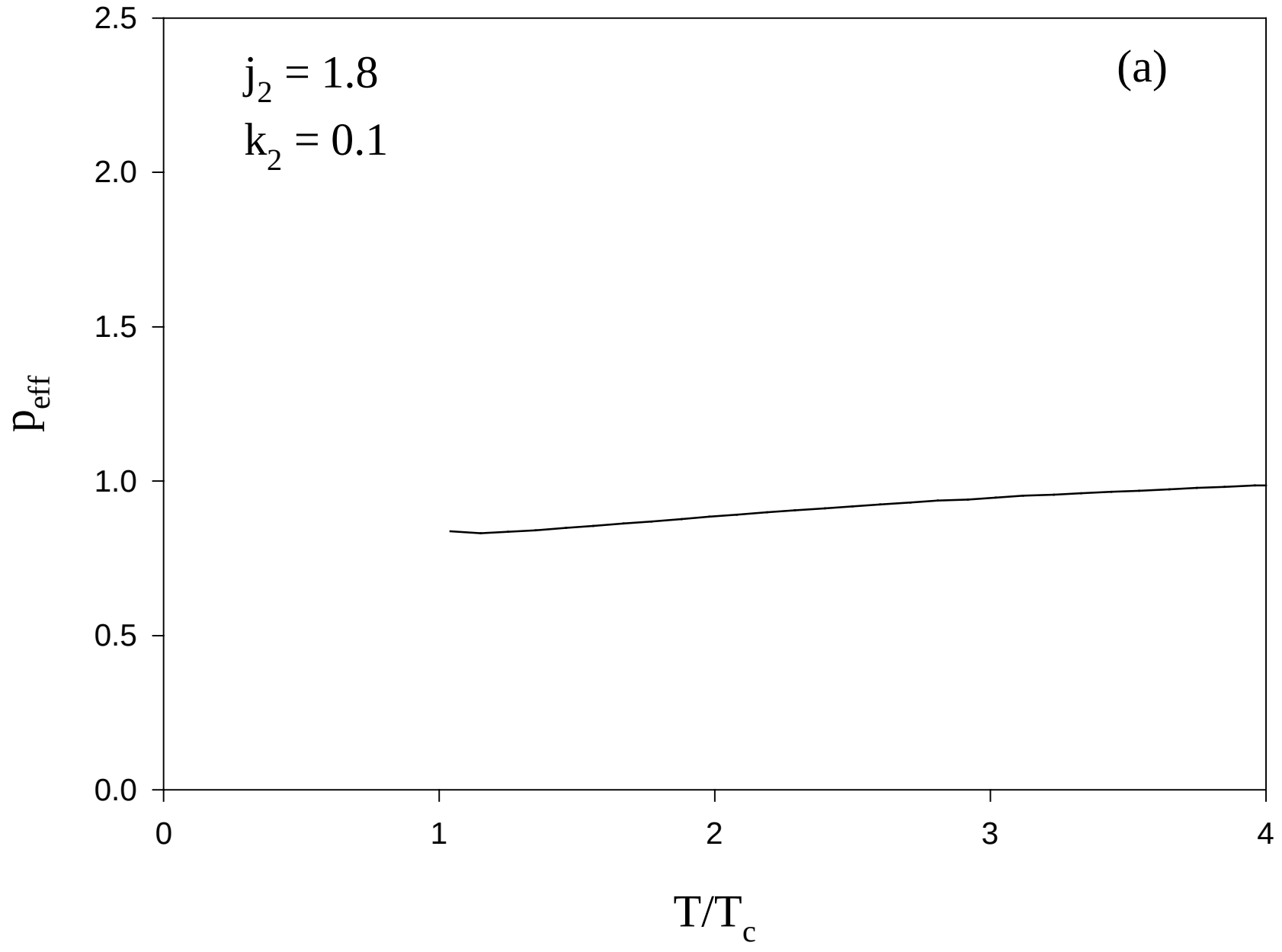

**Figure 7(a)**

Role of Inter-site Hybrid Interactions in Itinerant Ferromagnetism

Piyush Dua, Sunil Panwar and Ishwar Singh

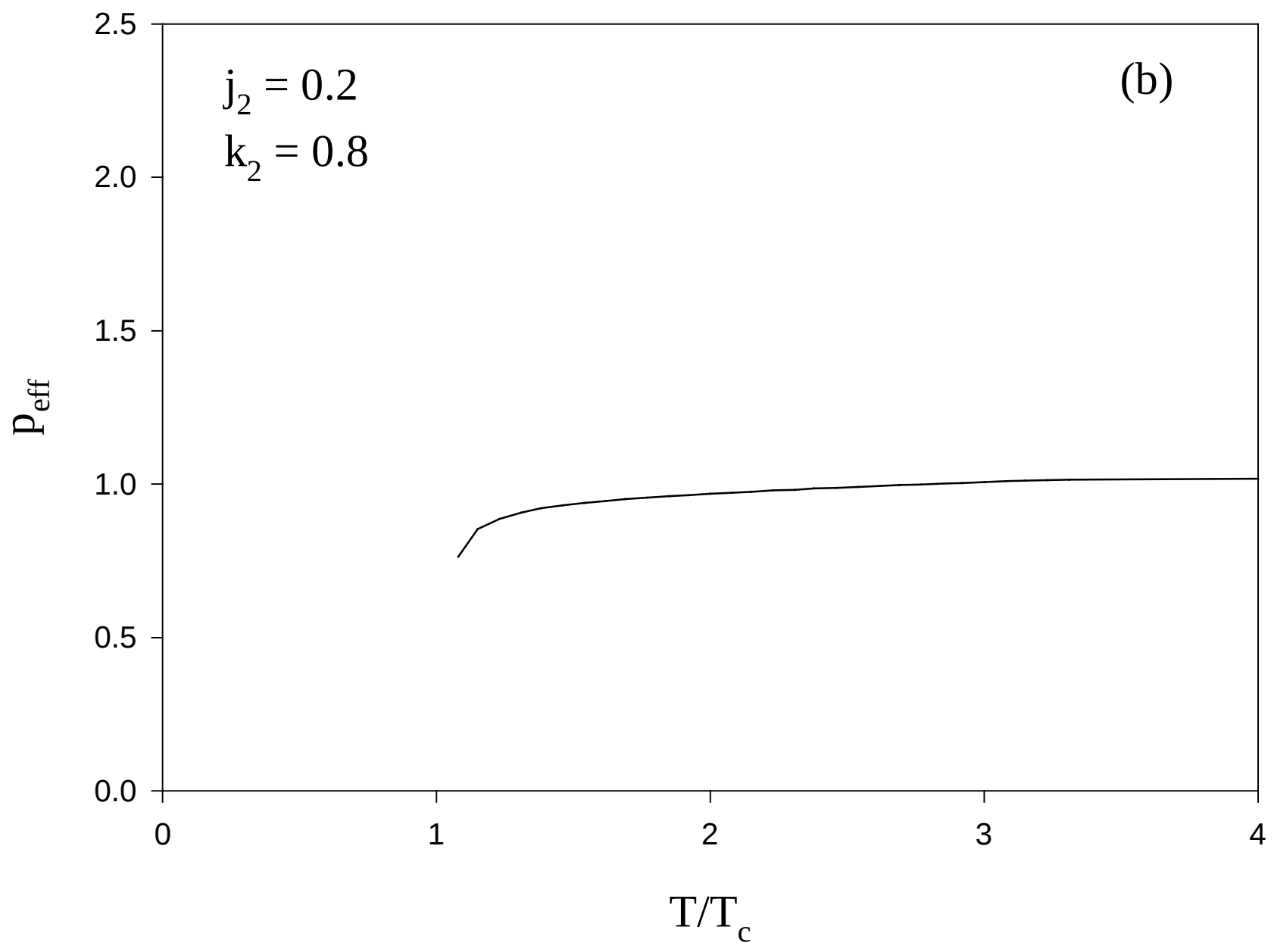

**Figure 7(b)**

Role of Inter-site Hybrid Interactions in Itinerant Ferromagnetism

Piyush Dua, Sunil Panwar and Ishwar Singh

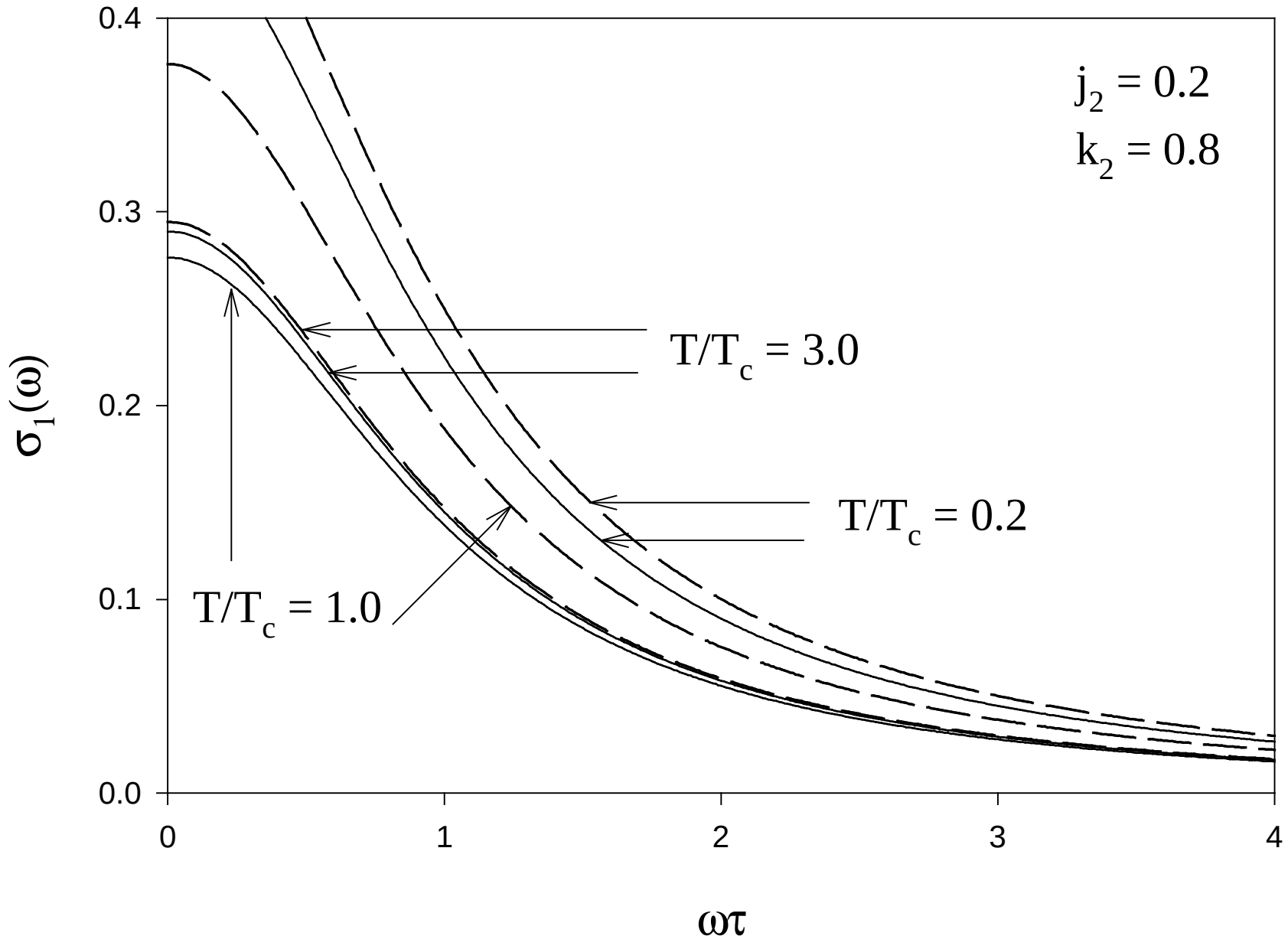

**Figure 8**

Role of Inter-site Hybrid Interactions in Itinerant Ferromagnetism

Piyush Dua, Sunil Panwar and Ishwar Singh

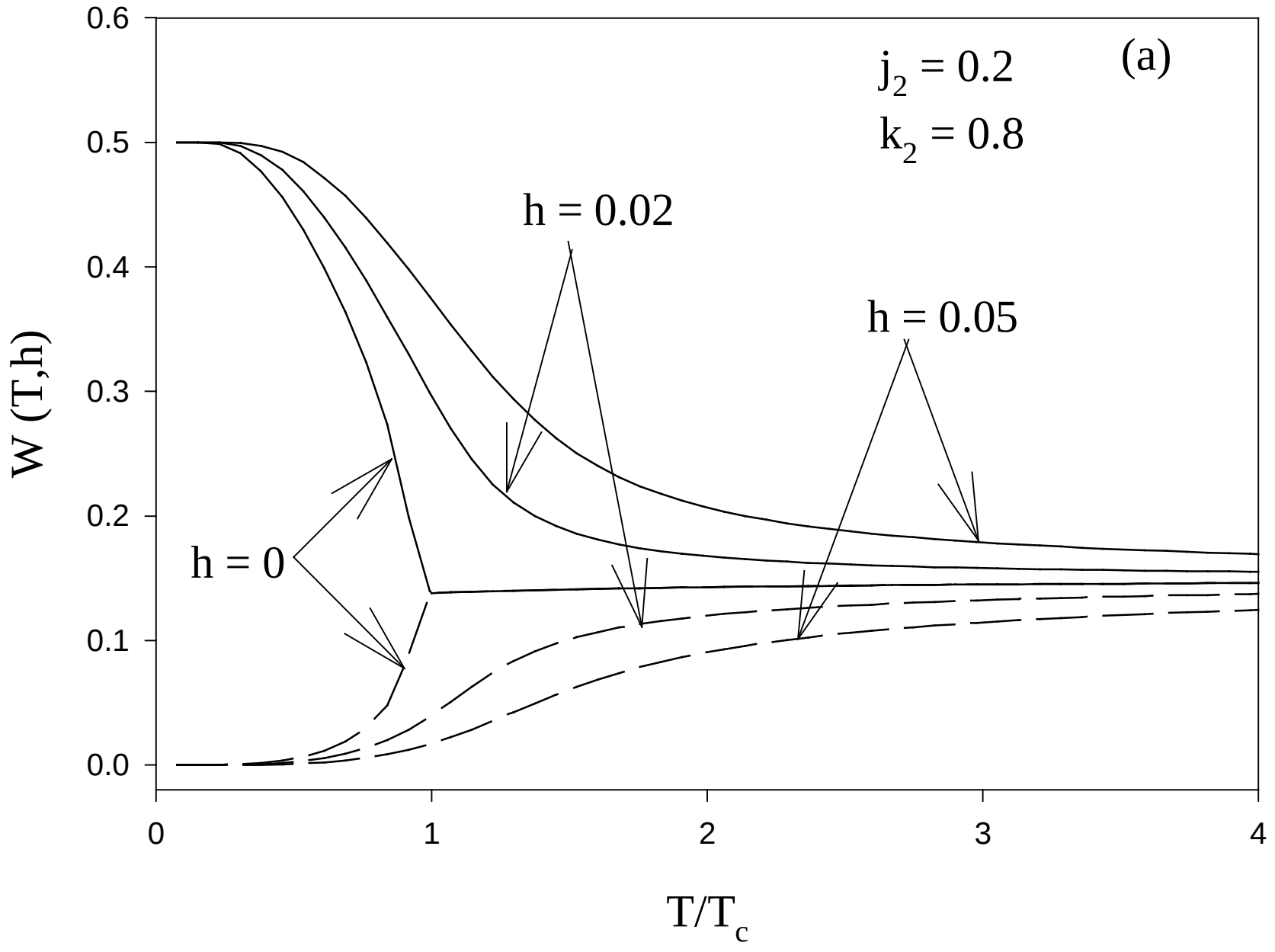

**Figure 9(a)**

Role of Inter-site Hybrid Interactions in Itinerant Ferromagnetism

Piyush Dua, Sunil Panwar and Ishwar Singh

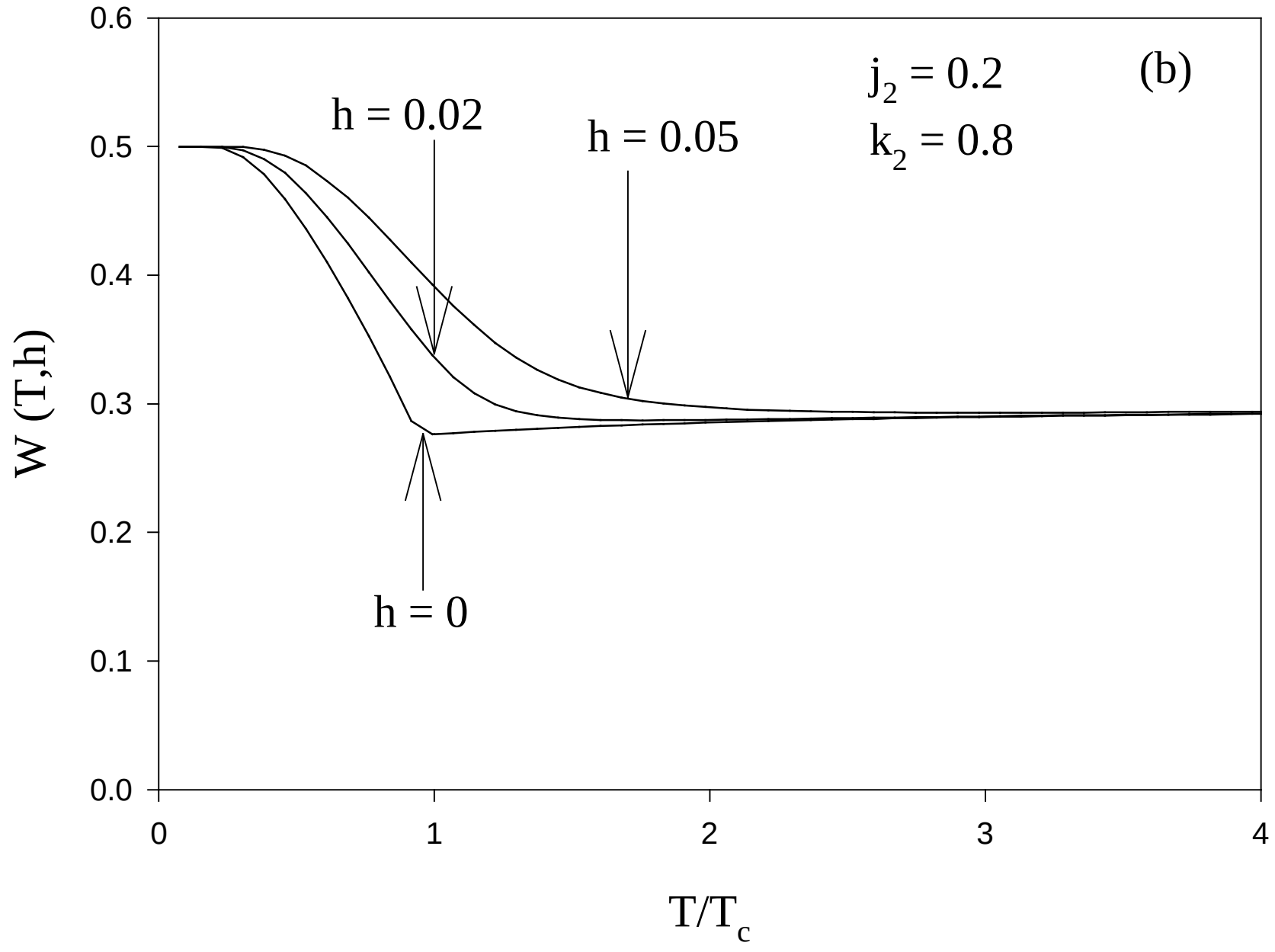

**Figure 9(b)**

Role of Inter-site Hybrid Interactions in Itinerant Ferromagnetism

Piyush Dua, Sunil Panwar and Ishwar Singh

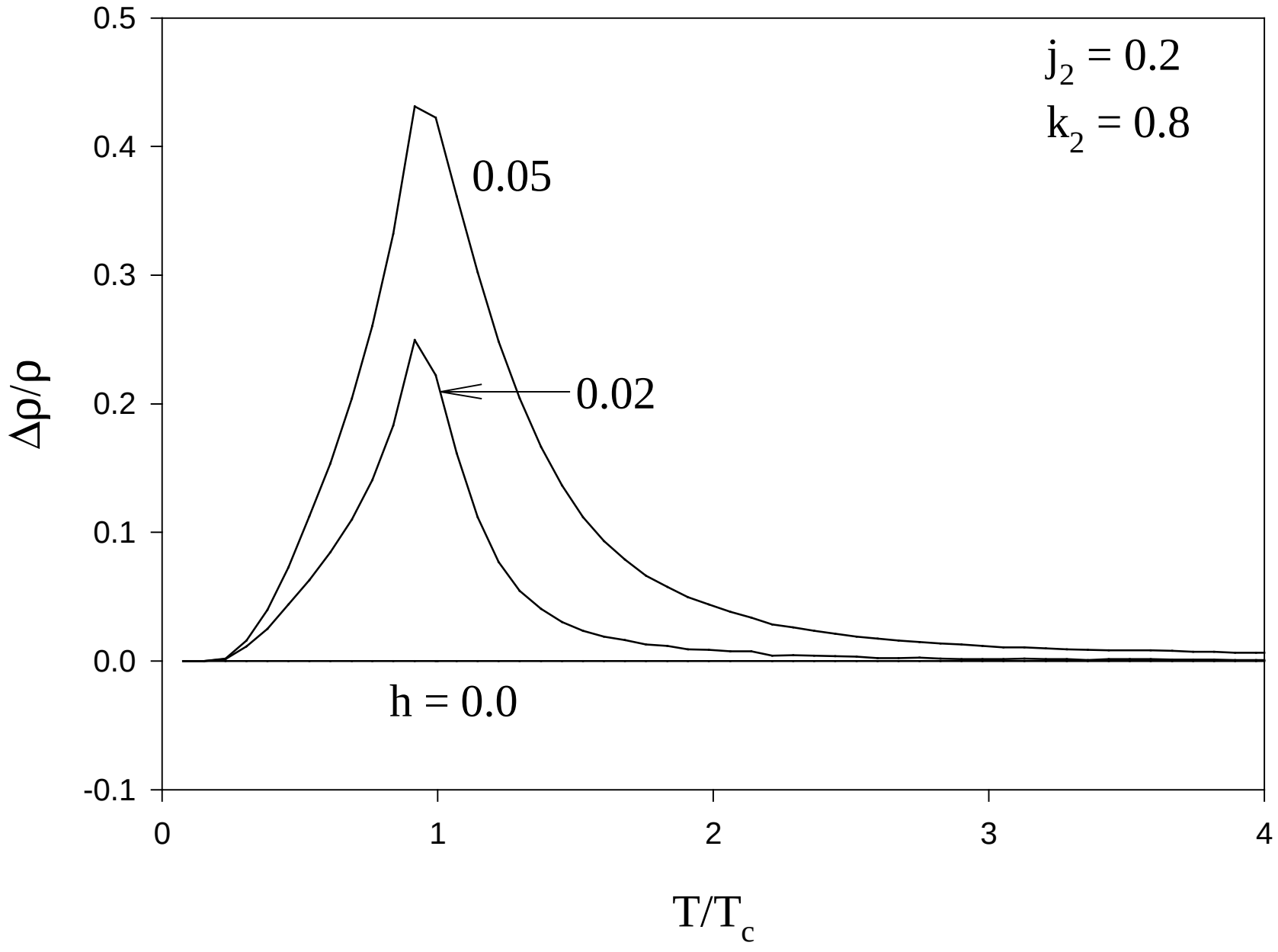

**Figure 10**

Role of Inter-site Hybrid Interactions in Itinerant Ferromagnetism

Piyush Dua, Sunil Panwar and Ishwar Singh